\documentclass[prl,showpacs,twocolumn]{revtex4}  
\usepackage{graphics}

\begin{document}

\title{Classical noise and flux: the limits of multi-state atom lasers}
\author{N.P. Robins}
\author{C. M. Savage}
\author{J. J. Hope}
\author{J. E. Lye}
\author{C. S. Fletcher}
\author{S. A. Haine}
\author{J. D. Close}

\affiliation{Australian Centre for Quantum Atom Optics, The Australian National University, Canberra, 0200, Australia.}
\email{nick.robins@anu.edu.au}
\homepage{http://www.acqao.org/}

\begin{abstract}
By direct comparison between experiment 
and theory, we show how the classical noise on a multi-state atom 
laser beam increases with increasing flux. The trade off between 
classical noise and flux is an important consideration in precision 
interferometric measurement. We use periodic $10 \mu$s radio-frequency pulses to couple atoms  out of an $F=2$ $^{87}$Rb Bose-Einstein condensate. The resulting atom laser beam has suprising structure which is explained using three dimensional simulations of the five state Gross-Pitaevskii equations.
\end{abstract}

\pacs{03.75.Pp,03.75.Mn}

\maketitle
It is the high flux, spectral density, and associated first order coherence that has made the optical laser central to many technologies. In the field of precision measurement, atom lasers hold similar promise \cite{ketterleandco}. In a Sagnac interferometer, for 
example, the inherent sensitivity of a matter wave gyroscope exceeds 
that of a photon gyroscope with the same particle flux and area by 11 orders 
of magnitude \cite{kasevich1}.  In any practical application of 
interferometry to high precision measurement, whether it be with 
photons or atoms, there will be a trade off between the classical 
noise and the quantum noise or flux of the source. In this Letter, we investigate this trade off for an atom laser.

In an unpumped laser, classical noise is the presence of unwanted excited dynamic modes.  We find  agreement between our experimental results and a full 3D Gross-Pitaevskii (GP) model, and show that at high flux, classical noise increases with increasing flux. That we can achieve agreement between a 3D theory including all Zeeman states and experiment is significant. It is highly likely that the much sought-after pumped atom laser will operate under rather specific 
conditions of scattering length, temperature and number density \cite{us}, and 
experiments will need to be guided by accurate theoretical models 
that must be validated against experiments if we are to trust their detailed predictions. 

Mewes {\it et al.} \cite{mewes} demonstrated the first atom laser based on the application of pulsed radio-frequency (RF) fields to induce controlled spin flips from magnetically trapped to un-trapped states of a Bose-Einstein condensate. Later it was shown by Hagley {\it et al.}  that a pulsed Raman out-coupling could be used to achieve a quasi-continuous multi-state atomic beam \cite{hagley}.   Bloch {\it et al.} achieved continuous RF out-coulping for up to 100 ms, producing a single state atom laser beam, and showed that this beam could be coherently manipulated in direct analogy to the optical laser \cite{bloch1, bloch2}. Both temporal and spatial coherence have been measured, and it has been demonstrated that RF outcoupling preserves the coherence of the condensate \cite{koehl,bloch3,kasevich}. The beam divergence has been measured \cite{aspect}, and there has been one real time measurement of the flux of an atom laser beam \cite{koehl2}. Ballagh {\it et al.} \cite{ballagh} introduced the Gross-Pitaevskii equation as an effective tool for investigating the atom laser within the semi-classical mean-field approximation and a number of groups found good agreement between GP models and experiment \cite{schneider,steck}. Rabi cycling between Zeeman components, a manifestation of the non-Markovian nature of the atom laser \cite{jack,hope}, was observed in the experiment of Mewes {\it et al.} and could be expected to significantly increase the amplitude, and possibly frequency, noise of the beam.  There has been no investigation of the relationship between classical noise and flux in an atom laser, and it is this aspect that we investigate both experimentally and theoretically in this Letter.

Experimentally, a continuous 
atom laser based on resonant output coupling puts stringent limits on the stability of cold atom traps \cite{bloch1}. 
Typical condensates have a resonant width of 10 kHz.  Output coupling requires a 
stable magnetic bias, $B_0$, at the 0.1 mG level, one to two orders of 
magnitude better than typical magnetic traps.  In comparison, a 
pulsed atom laser is relatively straight-forward to implement. A 
$10 \mu$s pulse has a frequency width of 200 kHz, significantly 
broader than both the 10 kHz resonant width of the condensate, and 
the instability of our trap which fluctuates within the range $\pm 15$~kHz due mainly to thermal fluctuations of the coils.  In the work 
reported in this Letter, we have opted to study a pulsed atom laser 
to ensure shot to shot reproducibility and allow detailed 
quantitative comparison to numerical models. We have chosen to study atom laser 
beams derived from an $F=2,m_F=2$ condensate because of the 
richness and complexity offered by the system, although non-Markovian effects are present even in two-component atom lasers due to the nonlinear atomic dispersion relations.

\begin{figure}[b]
\centerline{\scalebox{.45}{\includegraphics{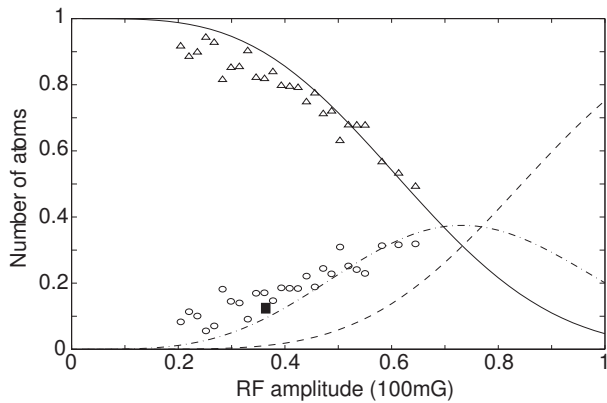}}}
\caption{Outcoupled fraction as a function of RF amplitude.  The solid square is the power at which the results of Fig.~\ref{experiment} were generated.  Theoretical curves are: Solid line, $m_F=2,1$ trapped states, dot-dashed $m_F=0$ and dashed, $m_F=-2,-1$ anti-trapped states.  The experimental results are: triangles, $m_F=2,1$ trapped states, circles  $m_F=0$. Typical error bars are $\pm 5\%$ vertically and $\pm 10\%$ horizontally.
Simulation parameter: $\Delta = -640$.}
\end{figure}
In our experiment, we 
produce an $F=2, m_{F}=2$  $^{87}$Rb condensate, consisting of 
approximately 50,000 atoms, via evaporation in a water-cooled QUIC 
magnetic trap \cite{esslinger} with a radial trapping frequency 
$\nu_r=253$ Hz, an axial trapping frequency $\nu_z=20$ Hz, and a 
bias field $B_0=1$ G. After evaporative cooling, the BEC was left 
to equilibriate for 100 ms. We then triggered an RF signal generator 
set in gated burst mode. The RF pulses were amplified (35 dB) and 
radiated perpendicular to the magnetic bias field of the trap through 
a 22 mm radius single loop, approximately 18 mm from the BEC.  To 
ensure that we had calibrated all experimental parameters correctly, 
we made an initial series of measurements of the number of trapped 
and un-trapped atoms after the application of single RF pulses of 
varying amplitude; it was critical to establish agreement between 
experiment and theory in a simple mode of operation before pursuing 
studies of more complex dynamics.  In Fig.~1, we show the results 
of these measurements in comparison with a {\em one dimensional} 
Gross-Pitaevskii theory of the atom laser derived from the full 3D model described by the following equations
\begin{equation}
 \label{secondset}
\begin{array}{l} 
{\displaystyle
i\dot\phi_2=({\mathcal{L}}+
V_T+Gy-2\Delta )\phi_2+2\Omega\phi_1}\\[7pt]
{\displaystyle
i\dot\phi_1=({\mathcal{L}}+
\frac{1}{2}V_T+Gy-\Delta )\phi_1+2\Omega\phi_2+\sqrt{6}\Omega\phi_0}\\[7pt]
{\displaystyle
i\dot\phi_0=({\mathcal{L}}+Gy) \phi_0+\sqrt{6}\Omega\phi_1+\sqrt{6}\Omega\phi_{-1}}\\[7pt]
{\displaystyle
i\dot\phi_{-1}=({\mathcal{L}}
-\frac{1}{2}V_T+Gy+\Delta )\phi_{-1}+2\Omega\phi_{-2}+\sqrt{6}\Omega\phi_0}\\[7pt]
{\displaystyle
i\dot\phi_{-2}=({\mathcal{L}}-
V_T+Gy+2\Delta )\phi_{-2}+2\Omega\phi_{-1} \,  ,}
\end{array}\end{equation}
where $\phi_i$ is the GP function for the $i$th Zeeman state. $V_T=\lambda^2(x^2+y^2)+z^2,\quad 
{\mathcal{L}}\equiv-\frac{1}{2}\nabla^{2}+U (\Sigma_{i=-2}^{2} |\phi_{i} |^2)$.  Here 
$\lambda=\nu_r / \nu_z = 12.65$ is the ratio of trapping 
frequencies, $\Delta$ and $\Omega$ are respectively the 
detuning of the RF field from resonance and the Rabi frequency, measured in units of $\omega_z =2 \pi \nu_z$. $U$ is the two-body interaction coefficient and $G= z_0 mg / (\hbar \omega_z)$, where $m$ is the atomic mass and $g$ the acceleration due to gravity. $z_0 = \sqrt{\hbar / m\omega_z }$ is the usual harmonic oscillator length. The wave 
functions, time, spatial coordinates, and interaction strengths are measured in the units of $z_0$, and $\omega_{z}^{-1}$. The excellent 
agreement between experiment and theory shown in Fig.~1 indicates 
that we have a well calibrated and repeatable experiment. Up to the experimental uncertainty in the detuning $\Delta$, there are no free parameters in the 1D GP model.  The theoretical results of the Rabi oscillations
presented in Fig.~1 are in good agreement with the approximate analytic theory presented by Graham and Walls for the limit of strong out-coupling \cite{graham}.  

In the experimental data shown 
in Fig.~2, we present five pulse trains outcoupled from separate 
F=2, $m_F =2$ condensates. In each case, the pulse train  has been 
outcoupled in an 8 ms time frame.  We wait 2 ms after the pulse 
window before turning the trap off to allow the final atomic pulse to 
completely separate from the condensate.  After a further 2 ms, to 
allow expansion of the condensate, we image the condensate and pulses 
with a single lens onto a 12 bit CCD camera.  For one, two, three and 
four RF pulses, we observe predictable out-coupling from the atom 
laser system.  Figure 2(a) is indicative of this behavior, where four 
RF pulses (separation 2 ms) have been applied to the BEC, and we see 
four $m_F=0$ atomic pulses in the positions expected from gravity. 
In 2(b) five RF pulses (separation 1.6 ms) have been applied and we 
observe five atomic wave-packets, again in the expected positions. 
However, we note that in the later 3 pulses there is a significant 
blurring with atoms appearing between the expected positions of the 
pulses.  This effect is not due to interference between the 
wavepackets.  

\begin{figure}[t]
\centerline{\scalebox{.45}{\includegraphics{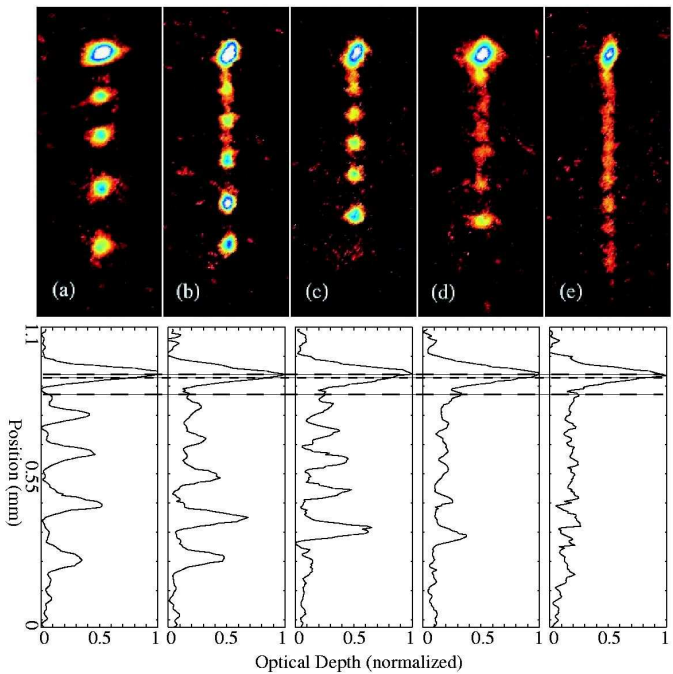}}}
\caption{(color) A series of pulsed atom lasers at different pulse 
rates.  The applied radio-frequency (RF) pulses are varied from (a) 4 
pulses, (b) 5 pulses, (c) 6 pulses, (d) 7 pulses, (e) 10 pulses in an 
8 ms window.  The lower plots show a cross-section down the centre of 
the absorption data.  The three dashed lines correspond in descending 
order to the centre of the condensate, the half-width of the 200 kHz 
RF resonance ($100$ kHz $\approx 12 \mu$ m in our trap) and the position 
coinciding with the final RF out-coupling pulse (4 ms prior to 
imaging). }
\label{experiment}
\end{figure}
The 
transition from constant pulse amplitude shown in Fig.~2(a) to 
varying or noisy pulse amplitude with increasing repetition rate shows a clear  trade off between 
classical noise and flux in the atom laser output. In Fig.~2(c) six RF pulses were applied (separation 
1.2 ms), however only five atomic pulses were observed, with the first 
atomic pulse being entirely absent.  This observation is quite 
repeatable. The complete 
absence of a pulse is an extreme example 
of the trade off between classical noise and flux, and it is the dynamics behind this 
phenomenon that we wished to understand by comparison with a 
complete 3D GP model.   At the higher pulse repetition rate used in 
Fig.~2(d) where the separation between pulses is 1 ms, the output 
is further distorted from the ideal. In Fig.~2(e), where the time 
between pulses has been reduced to $800 \mu$s, the atom laser beam is 
longer than expected from pure gravitational acceleration. This can 
be explained by the influence of the the anti-trapped $m_F$ states 
on the $m_F=0$ atoms that comprise the outcoupled beam. It is quite 
clear from the data that increasing flux (and therefore decreasing 
shot noise) comes at the price of increasing classical noise.  
 
We have quantitatively modelled the experiment with a full
3D GP simulation including all five Zeeman states and with only the detuning $\Delta$ as a free 
parameter  (Eq.~\ref{secondset}).    This is a unique feature of the work 
presented here and allows us to understand all aspects of the GP 
physics, and hence the experiment. A 1D model accurately describes a single out-coupling pulse because it is essentially independent of the spatial structure.  However we found that a full 3D simulation was needed to accurately track the spatio-temporal dynamics of a multi-pulse experiment. We 
simulated up to 3.2 ms, allowing 
three pulses for each case. Parallelised 
code was run on twelve processors of the APAC National Facility 
\cite{APACNF}, requiring up to 
800 hours of processor time per simulation. The numerical method was the psuedo-spectral method 
with Runge-Kutta split time step developed at the University of Otago 
\cite{RK4IP}. Spatial grid sizes and time steps were monitored 
throughout the simulations to ensure the accuracy of the numerical 
solutions; e.g. the preservation of the normalisation. Spatial 
grids were grown, and time steps decreased, as required. At the end of simulations, spatial grids in the direction 
of gravity were 2048 points for the trapped ($m_F=2,1$) and 
anti-trapped ($m_F=-2,-1$) Zeeman states, and 4096 for the  
atom laser output state ($m_F=0$). In the tight and loose transverse 
directions, 128 and 32 points were used respectively. The 
corresponding spatial lengths were chosen so that both the momentum 
space and real space GP functions fit the grid.  
This was about 40 $\mu$m and 140 $\mu$m in the tight and loose trap 
directions, respectively. In the direction of gravity, it was 120 
$\mu$m for the $m_F \neq 0$ states, and twice that 
for the $m_F = 0$ state. Absorbing boundaries were used for the $m_F=0,-1,-2$ states.

\begin{figure}[t]
\centerline{ \scalebox{.38} {\includegraphics{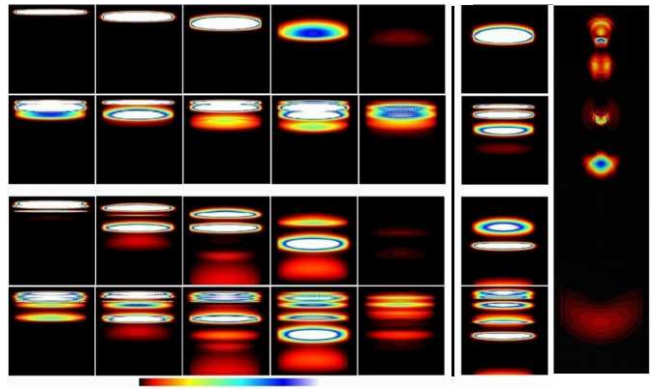} } }
\caption{(color)  Numerical simulations of the cases of Fig. 2(c) and 2(b). 
Case 2(c) is to the left of vertical line. Each image 
shows the GP wavefunction density, in arbitrary units,  integrated 
through the tight trap direction. Density is indicated by the colobar 
at the bottom of the figure, with black and white corresponding to 
zero and high density respectively. Each image is $120 \mu$m in both  
directions. From top to bottom the rows are: $t=1.2$ ms, just before 
the 2nd RF pulse; just after the 2nd RF pulse; $t=2.4$ ms, just 
before the 3rd RF pulse; just after the 3rd RF pulse. $t=0$ is the beginning of the first RF pulse. The columns 
from left to right are the $m_F=2,1,0,-1,-2$ states. The top-left image therefore shows the trapped position of the initial $m_F=2$ condensate. 
Case 2(b) is to the right of vertical line. Only the $m_F=0$ state is shown.
The rows are as before but the RF pulses occur at $t=1.6$ ms and $t=3.2$ ms.
Each image is $120 \mu$m vertically and $140 \mu$m horizontally. The rightmost image shows the $m_F=0$ state density on a slice plane through the tight trap direction, just after the 3rd RF pulse. It is $40 \mu$m horizontally and about $150 \mu$m vertically, allowing the first pulse to be seen.
Simulation parameters: $\Delta = -633$, $\Omega = 457$. See text for discussion.}
\label{simulation}
\end{figure}

The simulations reveal that {\em all five Zeeman states are involved in 
determining the final form of the atom laser output}. The $m_F=-2$ 
state has the least effect, as it is not strongly 
populated, and it quickly disperses in its anti-trapping potential. 
However the $m_F=-1$ anti-trapped state is highly populated and is 
directly involved in the loss of the initial pulse for the case of Fig.~2(c). 
The simulation in the left section of Fig.~\ref{simulation} shows how the first $m_F=0$ atom laser pulse is destroyed by the second 
RF pulse: it transfers nearly all of the 
$m_F=0$ component, produced by the first pulse, into the other four 
Zeeman states (second row, Fig.~\ref{simulation}).   
A new $m_F=0$ pulse, somewhat lower than the first, originates from the 
$m_F=-1$ state. However, it retains the  
momentum spread due to the anti-trapping potential, which causes it to 
disperse and be lost, so that it is not observed in the experiment.

The second atom laser pulse is in fact two distinct pulses; an upper one originating from the $m_F=2$ state, and a lower one from the $m_F=1$ state. This can be seen most clearly in the third row of  Fig.~\ref{simulation}, after they have become well separated.  Since the two pulse components are not resolvable in the experiment, this is an example 
of the dynamics revealed by simulation. 
These components have different initial momenta. The $m_F=1$ component, which originated from 
the $m_F=2$ state in the first RF pulse, was moving  down towards its 
trap equilibrium when the second RF pulse arrived.

The lower $m_F=0$ pulse created by the second 
RF pulse escapes the fate of the first pulse because its downward momentum takes it lower than the first pulse, away from resonance. In fact, its position is close to that of the 
upper second $m_F=0$ pulse during the third RF pulse in the case of Fig.~2(b), and it survives for similar reasons. This can be seen by comparing the bottom rows of the $m_F=0$ columns of Fig.~\ref{simulation}.  Similarly the simulations explain the relative intensity of the first 
and second $m_F=0$ pulses in the experimental case of Fig.~2(b). 

We have shown, for the case of the $F=2$ atom laser, that beyond a critical flux the 
classical noise on the output beam increases with increasing flux. 
The prospect of combining atom lasers with atom chips opens up 
enormous possibilities in precision measurement. Considerations of 
the trade off between classical noise and flux in atom lasers  will 
be important in many applications in this field. We would expect many 
of the effects described here to be smaller for the $F=1$ atom laser 
but not absent. Rather than the two trapped states present in the $F=2$ 
laser, only the $m_F=-1$ state is trapped. Although atoms in the 
$m_F=1$ state are anti-trapped for the $F=1$ laser, 
this state would be significantly populated for strong outcoupling 
and could be expected to contribute to classical noise on the $m_F=0$ 
output beam just as the anti-trapped $m_F=-1$ state does for the $F=2$ laser studied in this 
Letter. The effects that we have described will be important not only 
for pulsed atom lasers, but also for unpumped continuous atom lasers. 
Just as in the pulsed case, at high flux, atoms will not only be 
coupled to the output beam, they will also be coupled to other 
trapped and untrapped Zeeman states and can be backcoupled from the 
output beam to the condensate. The situation is complex and requires 
detailed investigation. The quantitative comparison between theory 
and experiment presented here is unique and points the way to the 
future development of atom laser sources for precision measurement. 
This is particularly true for the development of the pumped atom 
laser, one of the most important and sought-after devices in the field 
of atom optics. 

\acknowledgments
{This research was supported by the  Australian Partnership for Advanced Computing. ACQAO is an  Australian Research Council Centre of Excellence.}

\end{document}